\documentclass[usenatbib]{mn2e}
\usepackage{graphicx,ifthen,url}
\def\ltsima{$\; \buildrel < \over \sim \;$}
\def\lta{\lower.5ex\hbox{\ltsima}}
\def\gtsima{$\; \buildrel > \over \sim \;$}
\def\simgt{\lower.5ex\hbox{\gtsima}}

\newcommand {\um}{{\rm $\mu$m}}
\newcommand {\kms}{km\,s$^{-1}$}

\newcommand {\uJy}{{\rm $\mu$Jy}}
\newcommand {\Mpy}{M$_{\sun}$ yr$^{-1}$}
\newcommand {\aj}{AJ}
\newcommand {\aap}{A\&A}
\newcommand {\apj}{ApJ}
\newcommand {\apjs}{ApJS}
\newcommand {\apjl}{ApJL}
\newcommand {\mnras}{MNRAS}

\newcommand {\pasj}{PASJ}
\newcommand {\physrep}{Phys. Rep.}

\newcommand {\araa}{ARA\&A}
\newcommand {\qjras}{QJRAS}

\begin{document}
\loadboldmathitalic
\title[Hot-Dust ULIRGs at $z\sim1-3$]{
Confirming a Population of Hot Dust Dominant, Star Forming Ultraluminous Galaxies at High-Redshift}
\author[C.M. Casey et al.]{C.M. Casey$^{1}$\thanks{ccasey@ast.cam.ac.uk}, S.C. Chapman$^{1}$, 
R. Beswick$^{2}$, A.D. Biggs$^{3}$, A.W. Blain$^{4}$, 
L.J. Hainline$^{5}$, \newauthor 
R.~J. Ivison$^{6,7}$, T. Muxlow$^{2}$, Ian Smail$^{8}$ \\
$^{1}$ Institute of Astronomy, University of Cambridge, Madingley Road, 
Cambridge, CB3 0HA, U.K.\\
$^{2}$Jodrell Bank Centre for Astrophysics, University of Manchester, Oxford Road, Manchester, M13 9PL, U.K. \\
$^{3}$European Southern Observatory, Garching, Germany \\
$^{4}$Department of Astronomy, California Institute of Technology, 
1200 E California Blvd, Pasadena, CA, 91125, U.S.A. \\
$^{5}$Department of Astronomy, University of Maryland, College Park, MD, 20742, U.S.A.\\
$^{6}$UK Astronomy Technology Centre, Royal Observatory, Blackford 
Hill, Edinburgh, EH9 3HJ, U.K. \\
$^{7}$Institute for Astronomy, University of Edinburgh, Blackford Hill,
     Edinburgh, EH9 3HJ, U.K.\\
$^{8}$Institute for Computational Cosmology, Durham University, South 
Road, Durham DH1 3LE, U.K. \\
}

\date{Accepted 2009 June 19.  Received 2009 June 19; in original form 2009 May 26.}
\pagerange{\pageref{firstpage}--\pageref{lastpage}} \pubyear{2009}

\maketitle
\label{firstpage}

\begin{abstract}
We identify eight $z>1$ radio sources undetected at 850\um\ but robustly detected at 
70\um, confirming that they represent ultraluminous infrared galaxies (ULIRGs) with 
hotter dust temperatures ($<$T$_d>$=52$\pm$10\,K) than Submillimetre Galaxies (SMGs) 
at similar luminosities and redshifts.  These galaxies share many properties with 
SMGs: ultra-violet (UV) spectra consistent with starbursts, high stellar masses and 
radio luminosities.  We can attribute their radio emission to star formation since 
high-resolution MERLIN radio maps show extended emission regions (with half light
radii of 2-3\,kpc), which are unlikely to be generated by AGN activity.  These observations 
provide the first direct confirmation of hot, dusty ULIRGs which are missed by current 
submillimetre surveys. They have significant implications for future observations from 
the {\it Herschel Space Observatory} and {\sc SCUBA2}, which will select high redshift
luminous galaxies with less selection biases.
\end{abstract}

\section{Introduction}
The most luminous starbursts in the Universe signal a rapid growth phase in galaxies.
Submillimetre Galaxies \citep[bright infrared luminous galaxies with S$_{\rm 850}\simgt2$\,mJy at 
$<z>=2.2$; e.g.][]{smail02a,blain02a,chapman05a} represent some of the best studied ultra-luminous 
starbursts at high-$z$, and they are likely building some of the most massive galaxies found at the 
present epoch.  However, it has been suggested that selecting $z=$1-3 galaxies by their submillimeter 
emission is prone to finding specimens with cold average dust temperatures \citep[e.g.][]{eales00a,
blain04b,chapman04a} leaving the possibility of a large population of hotter ULIRGs at high
redshift.

The far-infrared (FIR) part of the spectrum (8-1000\um) is dominated by the re-processed emission from dust and 
resembles a modified blackbody whose most important variable is the characteristic dust temperature, 
T$_d$ \citep{blain02a}.  At $z\sim$2, the mean redshift of SMGs, submillimetre observations at 
850\um\ sample the emission at a strongly sloped section of the blackbody, the Rayleigh-Jeans tail, 
where the flux is approximated as S$_{\rm 850}\approx T_d^{-3.5}$ for a given $L_{\rm FIR}$.  For 
example, if T$_d$ rises (for fixed L$_{\rm FIR}$), the peak flux shifts towards shorter wavelengths, 
away from observed 850\um\ and the submillimeter (submm) flux can easily drop beneath the submm confusion limit 
\citep[$\sim$2\,mJy for JCMT;][]{blain99b}.  Current submm instruments restrict studies to those 
ULIRGs with cooler dust temperatures (T$_d<$40\,K) or the highest luminosities (L$>$10$^{13}$L$_\odot$),
potentially missing a substantial portion of the luminous starbursts at high redshift.

Observations of local ULIRGs \citep[e.g. $IRAS$ 60\um\ starbursts;
][]{levine98a} suggest that dust temperatures can vary substantially in star 
forming systems, from 20K up to 100K.  For a 10$^{13}$L$_\odot$ local ULIRG, 
the mean dust temperature is 45K \citep{rieke09a} higher than 36K, the 
mean temperature of similar luminosity SMGs at $z\sim$1-3.  A handful of individual 
ULIRGs at high redshift have been found to have characteristic dust temperatures 
around 50-80K (e.g. FSC\,10214+4724, Cloverleaf Quasar, IRAS\,F15307+3252, APM08297+5255). 
However, they are either exceptionally luminous or strongly lensed, thus detectable at 
70\um, 850\um\ and 1200\um\ wavelengths with the sensitivity of current instruments. This 
hints that submm selection may indeed be biased towards colder dust specimens, missing 
some fraction of hotter dust ULIRGs at high redshift.

At $z\sim$\,1-3, many galaxies with hot dust could exist but are difficult 
to detect and study as a population.  The first observational effort to 
characterize these ``hotter ULIRGs'' came in \citet{chapman04a}, with a selection 
of submm-faint radio-selected galaxies with starburst optical/near-IR spectra, 
similar to SMGs.  Since they had similar radio luminosities, redshifts, and 
optical spectra to SMGs they were thought to be ULIRGs.  However, poor constraints on 
their FIR luminosities provided insufficient evidence of their high bolometric luminosity.  

We present eight examples of ULIRGs where observations suggest hot dominant dust 
temperatures. \S \ref{sec:data} describes the data and our observations, \S \ref{sec:results}
discusses the galaxies' physical properties and \S \ref{sec:discussion} discusses the
selection biases prominent at FIR wavelengths and the context of hot-dust ULIRGs
in other $z\sim2$ galaxy populations.  Throughout, we use $AB$ magnitudes and assume 
H$_0$=\,71\,\kms\,Mpc$^{-1}$ and $\Omega_0$=0.27 \citep{hinshaw09a}.

\section{Sample and Observations}\label{sec:data}

\begin{table*}
\caption{Observed and derived properties of hot-dust ULIRGs.}
\label{tab1}
\begin{tabular}{c@{ }c@{ }c@{ }c@{ }c@{ }c@{ }c@{ }c@{ }c@{ }c@{ }c@{ }c@{ }c}
\hline\hline
NAME & $z$ & S$_{\rm 1.4\,GHz}$ & S$_{\rm 70}$ & S$_{\rm 850}$ & S$_{\rm 1200}$ & S$_{\rm 24}$ & L$_{\rm X[2-10\,keV]}$ & L$_{\rm FIR}$    & SFR                 & M$_\star$   & R$_{\rm eff}$ & T$_d$ \\
     &     & (\uJy)      & (mJy)   & (mJy)        & (mJy)           & (\uJy)   & (erg\,s$^{-1}$) &(L$_\odot$) & (M$_\odot$\,yr$^{-1}$) & (M$_\odot$) & (kpc) & (K)   \\
\hline
RG\,J123710.60+622234.6 & 1.522 & 38.3$\pm$10.1  & 3.9$\pm$0.5 & $<$1.8 & $<$1.8 & 227$\pm$39 & $<$9.6$\times$10$^{42}$ & 1.4$\times$10$^{12}$ & 247$^{+75}_{-57}$ & 1.2$\times$10$^{11}$ & 1.7 & 54$\pm$3 \\
RG\,J123653.37+621139.6 & 1.275 & 86.7$\pm$8.3   & 6.6$\pm$0.4 & $<$1.4  & $<$0.7 & 164$\pm$33 & 2.2$\times$10$^{42}$    & 2.0$\times$10$^{12}$ & 349$^{+35}_{-32}$ & 1.7$\times$10$^{11}$ & 2.9 & 47$\pm$1 \\
RG\,J123645.89+620754.1 & 1.433 & 83.4$\pm$9.8   & 4.8$\pm$0.4 & $<$6.3  & $<$1.6 & 172$\pm$34 & 5.1$\times$10$^{42}$    & 2.7$\times$10$^{12}$ & 458$^{+57}_{-52}$ & 3.0$\times$10$^{11}$ & 2.1 & 48$\pm$1 \\
RG\,J105159.90+571802.4 & 1.047 & 74.5$\pm$5.6   & 7.7$\pm$1.2 & $<$5.9  & $<$1.4 & 738$\pm$27 & $<$2.2$\times$10$^{42}$  & 1.0$\times$10$^{12}$ & 179$^{+14}_{-13}$ & 6.6$\times$10$^{10}$ & 2.7 & 46$\pm$2 \\
RG\,J105154.19+572414.6 & 0.922 & 45.4$\pm$6.3   & 6.1$\pm$0.9 & $<$4.3  & $<$1.7 & 510$\pm$22 & $<$1.6$\times$10$^{42}$  & 4.6$\times$10$^{11}$ & 79$^{+12}_{-10}$ & 8.3$\times$10$^{10}$ & 2.0 & 45$\pm$3 \\
RG\,J105146.61+572033.4 & 2.383 & 33.5$\pm$5.8   & 9.5$\pm$1.1 & $<$2.4 & $<$3.1 & 298$\pm$16 & $<$1.4$\times$10$^{43}$  & 4.1$\times$10$^{12}$ & 709$^{+120}_{-120}$ & 2.7$\times$10$^{11}$ & 2.1 & 72$\pm$3 \\
 {\bf ULIRGs with AGN:}                     &       &                &             & &   & & & & & & & \\
RG\,J123711.34+621331.0 & 1.996 & 126.3$\pm$8.6  & 1.4$\pm$0.4 & $<$6.0  & $<$4.4 & 473$\pm$57 & 1.9$\times$10$^{43}$    & 9.7$\times$10$^{12}$ & - & 3.5$\times$10$^{11}$ & 3.7 & 50$\pm$2 \\
$RG\,J123711\ SF$ component: &  & 79.6$\pm$17.2  &  &  & & & & 6.1$\times$10$^{12}$ & 1047$^{+120}_{-107}$ & & \\
RG\,J123649.66+620738.0 & 2.315 & 327.0$\pm$8.6 & 5.8$\pm$0.4 & $<$8.4  & $<$1.1 & 912$\pm$78 & 1.3$\times$10$^{45}$    & 3.7$\times$10$^{13}$ & - & 8.1$\times$10$^{10}$ & 2.7 & 57$\pm$1 \\
$RG\,J123649\ SF$ component: &  & 196.6$\pm$17.2  &  &  & & & & 2.2$\times$10$^{13}$ & 3843$^{+171}_{-164}$ & & \\
\hline\hline
\end{tabular}
{\small
Observed and derived properties of the eight submm-faint radio galaxies detected at 70\um.  
SFR and L$_{\rm FIR}$ are derived from radio luminosities, M$_\star$ from stellar population 
model fits, and T$_d$ from modified black body FIR continuum fits constrained by 70\um\ flux 
densities and normalized to radio luminosity.  The temperature uncertainties stated here do 
not take the uncertainty in the FIR/radio correlation into account (as described in section 
\ref{sec:dusttemperature}).  All sources are undetected at 850\um\ and 1200\um; 
the fluxes in the corresponding columns are 2$\sigma$ upper limits.  Two sources have AGN in 
addition to luminous star formation (RG\,J123711 and RG\,J123649); their L$_{\rm FIR}$ and SFR have 
been adjusted to represent just the star forming components in the abbreviated rows `RG\,J... SF component:'; 
the star forming components have been carefully deconvolved from AGN emission in the MERLIN maps 
as discussed in \S \ref{sec:agn}.  The effective radius, R$_{\rm eff}$, is the circularized radius 
corresponding to a surface area of the $>$3$\sigma$ MERLIN radio emitting region.
}
\end{table*}

We identified 44 potential hot high-$z$ ULIRGs using the \citet{chapman04a} 
optically faint radio galaxy (``OFRG'', also submm-faint, star forming radio galaxy, ``SFRG'') 
selection in the GOODS-N and Lockman Hole fields.  They were detected in the ultra-deep VLA radio maps of 
\citet*{biggs06a} (S$_{\rm 1.4\,GHz}>$15\uJy, $>$3\,$\sigma$).  Spectroscopic 
follow up of these potential ULIRGs with Keck LRIS revealed starburst spectral 
features \citep{chapman04a,reddy06a}, mostly at redshifts z\simgt1.  Obvious 
spectral AGN were not included in the sample.  There were 26 GOODS-N sources 
and 18 Lockman Hole galaxies selected by this radio/rest-UV spectra method.  
Their redshift distribution (quartile of $z\,=\,$1.2-2.6) is similar to 
the redshift distribution of radio-selected SMGs (quartile of $z\,=\,$1.2-2.8).

The $Spitzer$ 70\um\ maps of GOODS-N and Lockman Hole are used for further selection.  From the
sample of 44 radio-selected galaxies, five out of 26 in GOODS-N and three out of 18 are detected at 
70\um\ above $3\sigma$.  These eight galaxies make up the sample we discuss in this paper. The 
positional offsets between the 70\um\ centroids and the VLA radio positions are no more than 
5\arcsec, which is sufficiently smaller than the 70\um\ beam size and astrometric uncertainty.  
See Table~\ref{tab1} for a summary of the observational data.  We use the remaining 21 GOODS-N 70\um\ 
undetected galaxies in a stacking test described in section \ref{sec:undetected} (the Lockman 
sample is excluded for this test since its 70\um\ map has a much shallower depth).

All sources were observed with the Multi-Element Radio Linked Interferometer Network 
\citep[MERLIN;][]{thomasson86a,muxlow05a}, yielding high-resolution 0.3\arcsec\ 
(GOODS-N) and 0.5\arcsec\ (Lockman Hole) beam 1.4-GHz radio maps.  Three of the eight
galaxies were included in the \citet{muxlow05a} catalog of HDF radio sources (RG\,J123711, 
RG\,J123653, and RG\,J123645).  The 1200\um\ flux limits come from the Max-Planck Millimetre 
Bolometer \citep[MAMBO;][]{greve08a} and 850\um\ flux limits from SCUBA  \citep{borys03a,coppin06a}.  
Both fields are covered with $Spitzer$ IRAC (3.6, 4.5, 5.8, and 8.0\um) and MIPS (24 and 70\um), 
however at greater depths in all bands in GOODS-N. Optical photometry in GOODS-N is from the 
$HST$ ACS\footnote{Based on observations 
made with the NASA/ESA Hubble Space Telescope, and obtained from the Hubble Legacy Archive, 
which is a collaboration between the Space Telescope Science Institute (STScI/NASA), the 
Space Telescope European Coordinating Facility (STECF/ESA) and the Canadian Astronomy Data 
Centre (CADC/NRC/CSA).} using the F435W, F606W, F814W, and F850LP filters (B, V, i and z 
bands). The Lockman Hole has $HST$ ACS F814W only (PI: Chapman HST 7057).  One source in 
Lockman Hole (RG\,J105146) is not covered by ACS so we use Subaru/Suprime-Cam i-band instead 
\citep{miyazaki02a}.  The GOODS-N sources have X-ray coverage from the 
CDF-N.  All are formally undetected in the \citet{alexander03a} catalog above 5$\sigma$ in the 
full band (0.5-8.0\,keV), yet four sources are marginally detected at fluxes 
$<$6$\times$10$^{-16}$erg\,s$^{-1}$\,cm$^{-2}$.  Lockman Hole sources have X-ray coverage from 
$XMM$-$Newton$ \citep{brunner08a}, and none of the sources are formally detected, with a sensitivity
limit in the soft band (0.5-2.0\,keV) of 1.9$\times$10$^{-16}$\,erg\,cm$^{-2}$\,s$^{-1}$ and 
9$\times$10$^{-16}$\,erg\,cm$^{-2}$\,s$^{-1}$ in the hard band (2.0-10.0\,keV).  No 
X-ray signal is detectable in a stack of the Lockman Hole sources.

\section{Results}\label{sec:results}

\subsection{Dust Temperature}\label{sec:dusttemperature}

\begin{figure}
\centering
\includegraphics[width=0.99\columnwidth]{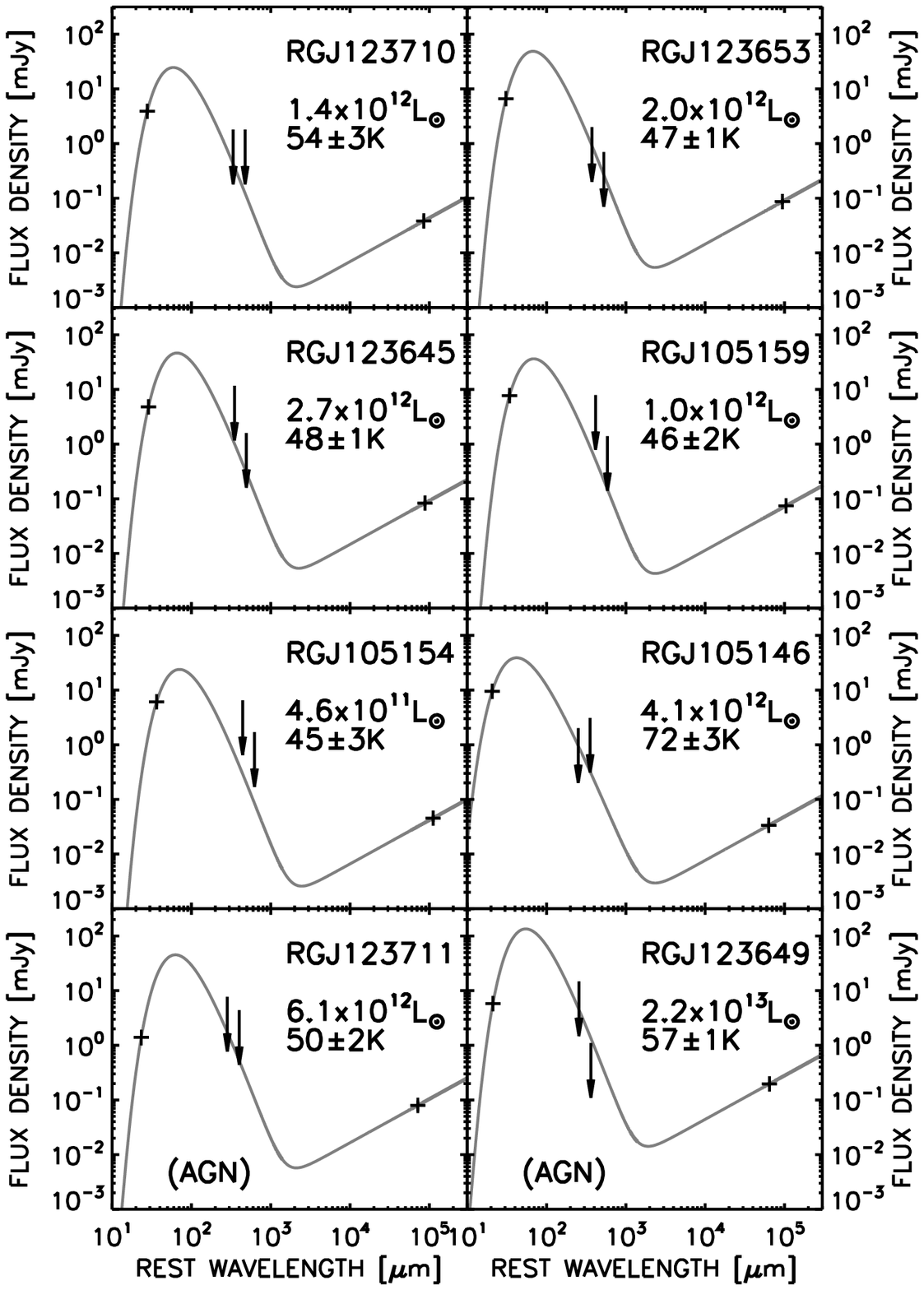}
\caption{FIR to radio spectral energy distributions of eight hot-dust ULIRGs. Each galaxy is fit 
with a modified blackbody SED to the 70\um\ flux density, limits at 850\um\ and 1200\um, and radio 
luminosity (which is taken to relate directly to $L_{\rm FIR}$ via the FIR/radio correlation).  
On this plot, there is no visible difference between $\beta=1.5$ and $\beta=2$ fits.  Inset on each
panel are the galaxy name, FIR luminosity (integrated from 8-1000\um) and best fit dust temperature.
The two panels marked with ``(AGN)'' show the SEDs for galaxies who show some evidence of containing 
luminous AGN (details described in section~\ref{sec:agn}).
}
\label{fig:all_seds}
\end{figure}

We fit FIR spectral energy distributions (SEDs) to the FIR flux densities (at 70\um, 850\um, and 1200\um)
using a modified blackbody emission curve where
\begin{equation}
\centering
S_\nu \propto \frac{\nu^{3+\beta}}{exp(h\nu/k T_d)-1}
\end{equation}
$S_\nu$ is the observed flux density at rest frequency $\nu$, $T_d$ is the characteristic
dust temperature and $\beta$ is the dust's emissivity.  Since we are limited by a lack
of data points in the FIR, we must assume that the FIR/radio correlation holds \citep[i.e. 
that $L_{\rm FIR}$, evaluated from 8-1000\um, scales linearly with 1.4GHz radio luminosity;][]{condon92a}.
While recent works \citep[e.g.][]{bell03a,beswick08a} indicate that the correlation might deviate 
at higher redshifts and lower flux densities, its effect on the dust temperature calculation would 
be systematic.  The 0.2\,dex scatter found for the local FIR-radio relation suggests an additional 
uncertainty to our $T_d$ estimates.  We note that the additional uncertainties would apply equally to the 
galaxies in this paper as well as the \citet{chapman05a} SMGs; foremost, we wish to highlight the temperature 
difference between the two populations rather than their absolute temperatures.

To calculate temperature, we require an additional constraint to reasonably fit 
the small number of photometric points with a modified blackbody. We fix $\beta=1.5$ and adopt a 
single temperature characterization of the emission, although it is recognized that dust components 
of a range in temperatures are required to accurately describe well studied nearby galaxies
\citep[e.g.][]{dale01a}.  We have tested that our fits are relatively insensitive to $\beta$ 
by also using $\beta=2$, and finding only small ($<2\,K$) differences in the fitted $T_d$.  
Fig.~\ref{fig:all_seds} shows the best fit modified blackbodies and derived temperatures for our galaxy sample.
These blackbody SEDs are combined with radio synchrotron emission ($\alpha=0.8$) which is fit through
the 1.4\,GHz flux densities.  Best fit dust temperatures are listed in Table~1.

\subsection{Stellar Mass}

\begin{figure}
\centering
\includegraphics[width=0.99\columnwidth]{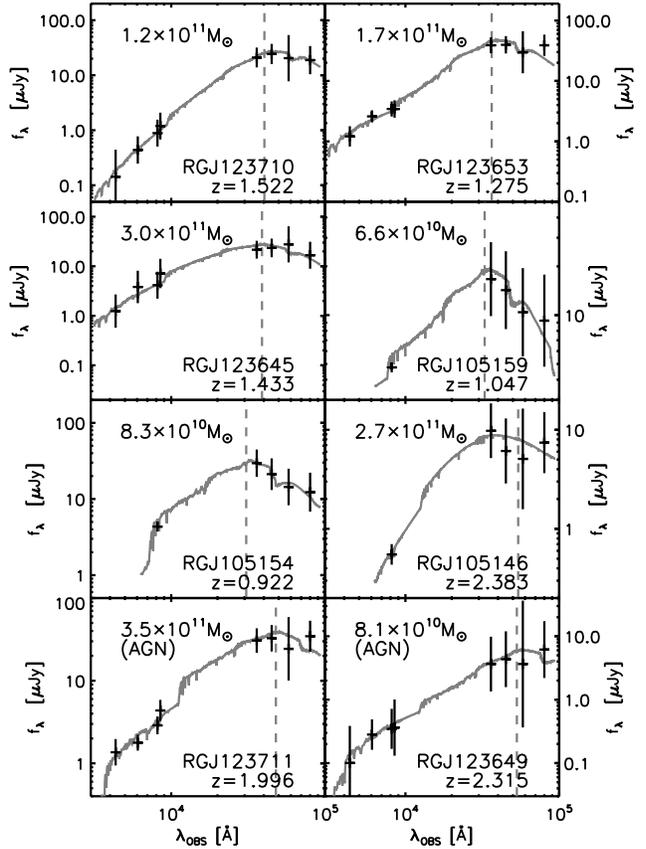}
\caption{
Optical and near-IR photometry points are fit with stellar population 
models.  From these SED fits, we interpolate rest-K band magnitudes 
and compute stellar masses using the method of \citet{borys05a}; the
derived stellar masses are given on each panel in the upper left. We also 
assess the AGN content based on an observed 8\um\ flux excess, which
is only statistically significant in the spectrum of RG\,J123711.  We 
mark the redshifted 1.6\um\ stellar bump with a dashed vertical 
line.  The two panels marked with ``(AGN)'' show the SEDs for galaxies 
who show some evidence of containing luminous AGN (details described in 
section~\ref{sec:agn}).
}
\label{fig:hyperzfits}
\end{figure}

To estimate the galaxies' stellar masses, we combine the photometric points in the optical 
($HST$ ACS $B$, $V$, $i$, and $z$ for GOODS-N and $HST$ ACS $i$ for Lockman Hole) and the 
near-IR ($Spitzer$-IRAC 3.6\um, 4.5\um, 5.8\um\ and 8.0\um).  For $z=$1-3, these photometric points
cover the rest-frame 1.6\um\ wavelength where stellar emission peaks.  We use the {\sc 
hyperz} photometric redshift code \citep{bolzonella00a} to fit this photometry to several 
stellar population SEDs \citep{coleman80a}.  The optimized stellar population fits are shown 
in Fig.~\ref{fig:hyperzfits}.  Measured output includes internal extinction factor A$_V$ and rest-frame 
K-band magnitude \citep[which we use to measure stellar mass with the method outlined by][assuming
a mean light-to-mass ratio of L$_K$/M = 3.2]{borys05a}. This analysis indicates that the 
systems are quite massive and comparable to the stellar masses of the Borys et~al. SMGs, with mean 
M$_\star \sim\,$2$\times$10$^{11}$M$_\odot$.  We note that \citet{alexander08a} and \citet{chapman09a} 
point out that the Borys et~al. method of using K-band luminosities overestimates stellar masses since 
it does not correct for the AGN contribution to 8\um\ flux density or sample the stellar emission at its
peak; while also a probable systematic error in our data set, the 
effect applies equally to SMGs and hot-dust ULIRGs.  Only one source, RG\,J123711, shows an excess 
8\um\ flux, indicative of AGN power law emission; its stellar mass is likely overestimated.  Derived 
stellar masses are listed in Table~1.

\subsection{MERLIN Radio Morphology}\label{sec:MERLIN}

\begin{figure}
\centering
\includegraphics[width=0.49\columnwidth]{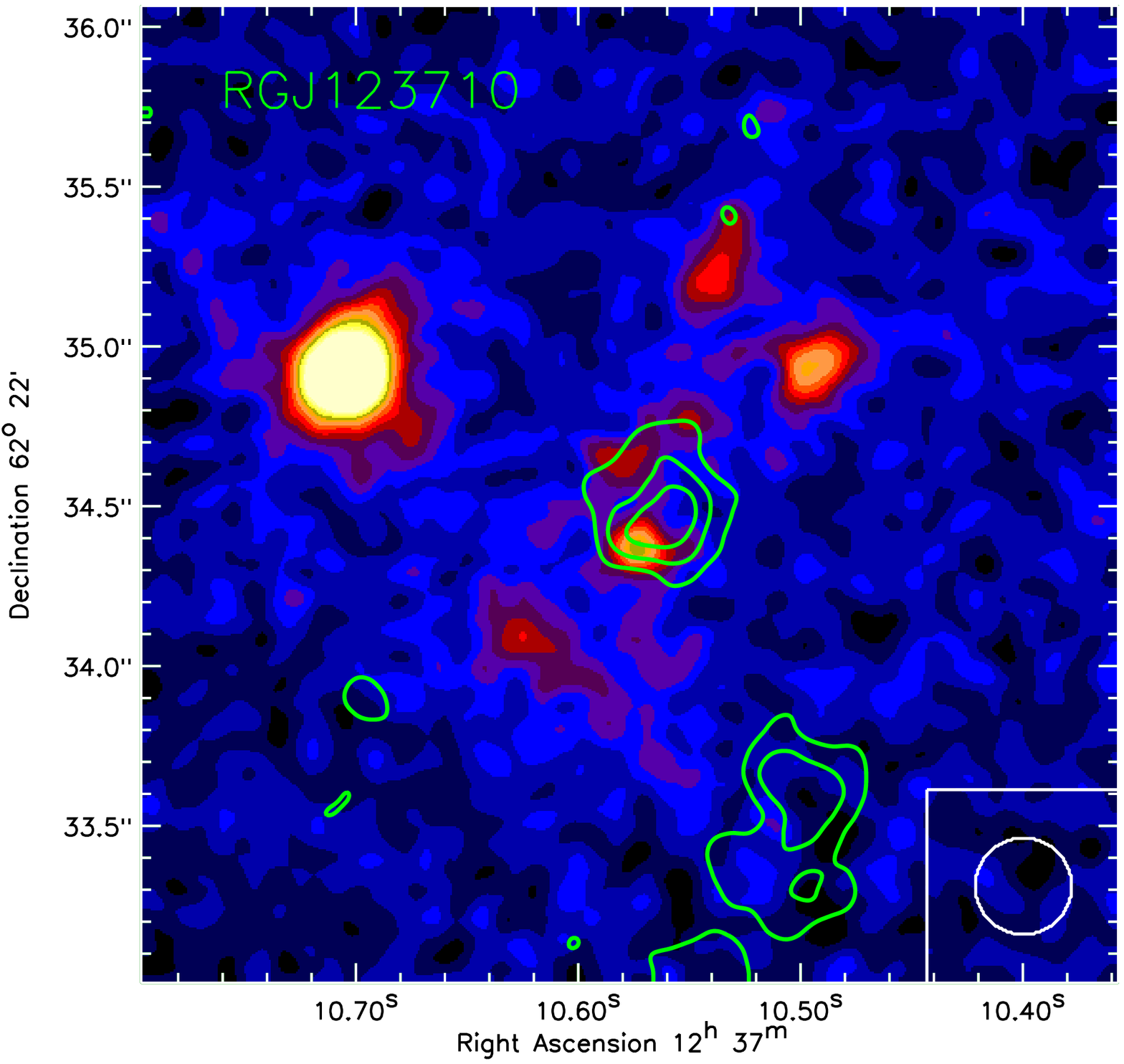}
\includegraphics[width=0.49\columnwidth]{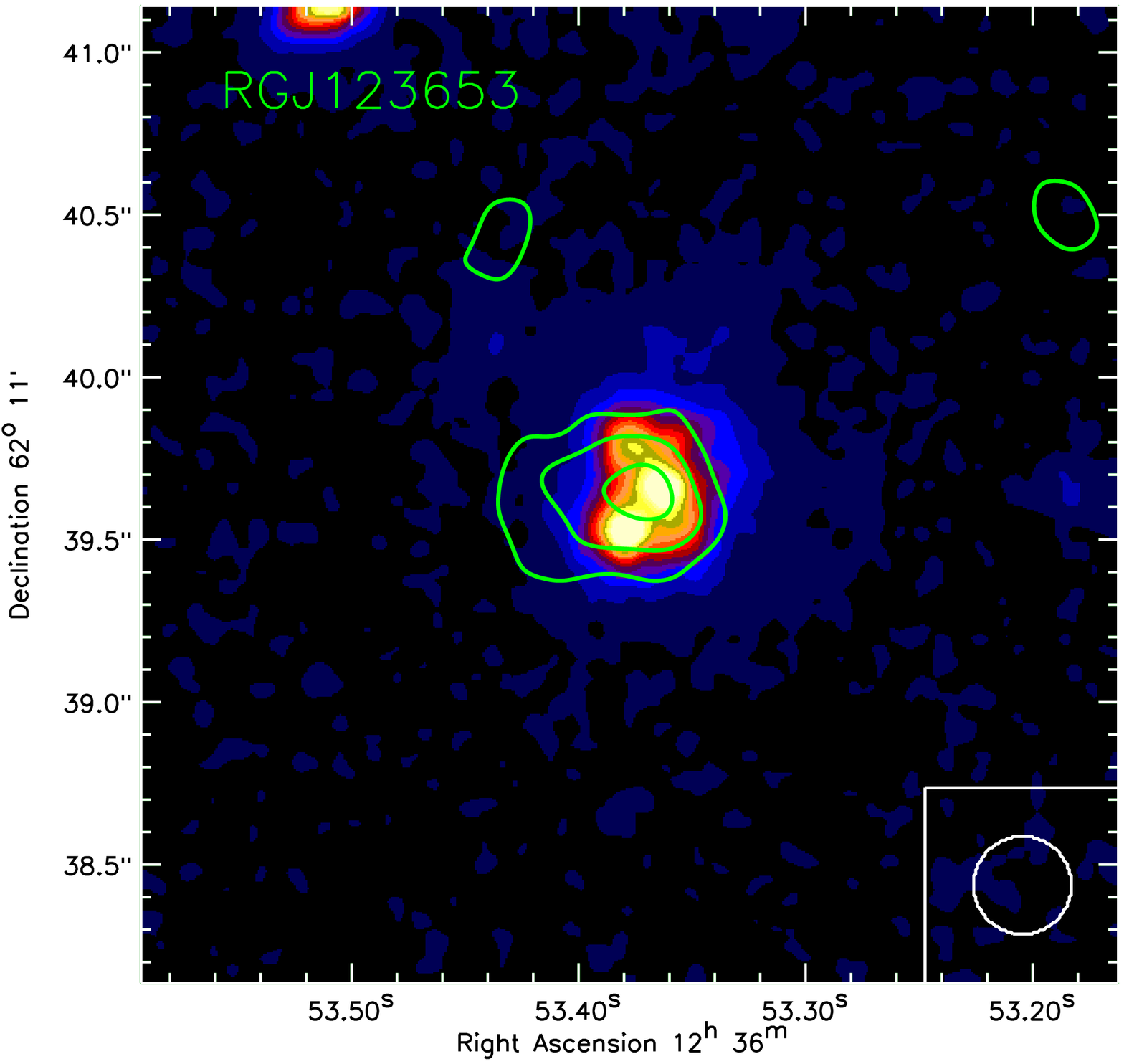}\\
\includegraphics[width=0.49\columnwidth]{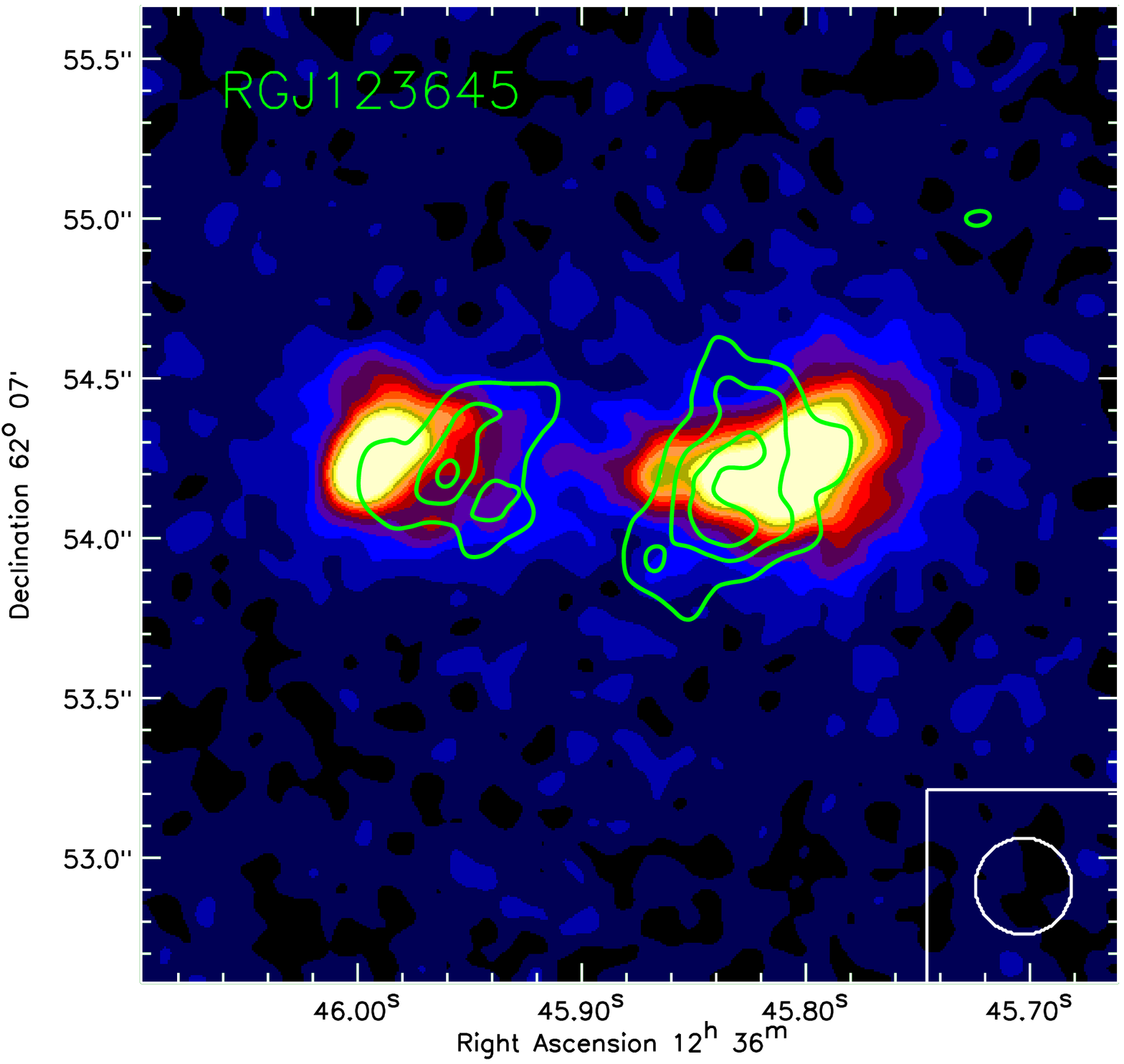}
\includegraphics[width=0.49\columnwidth]{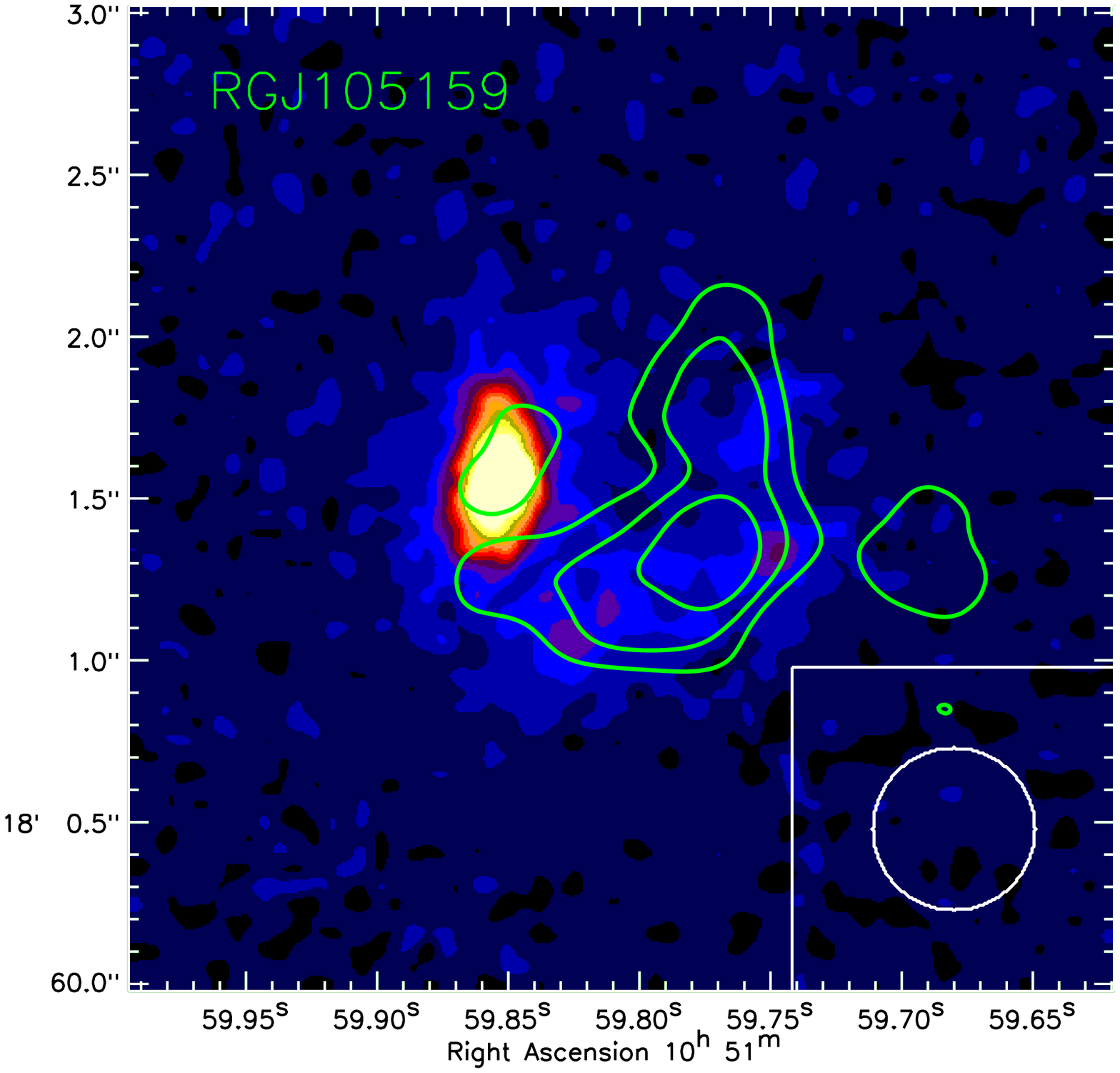}\\
\includegraphics[width=0.49\columnwidth]{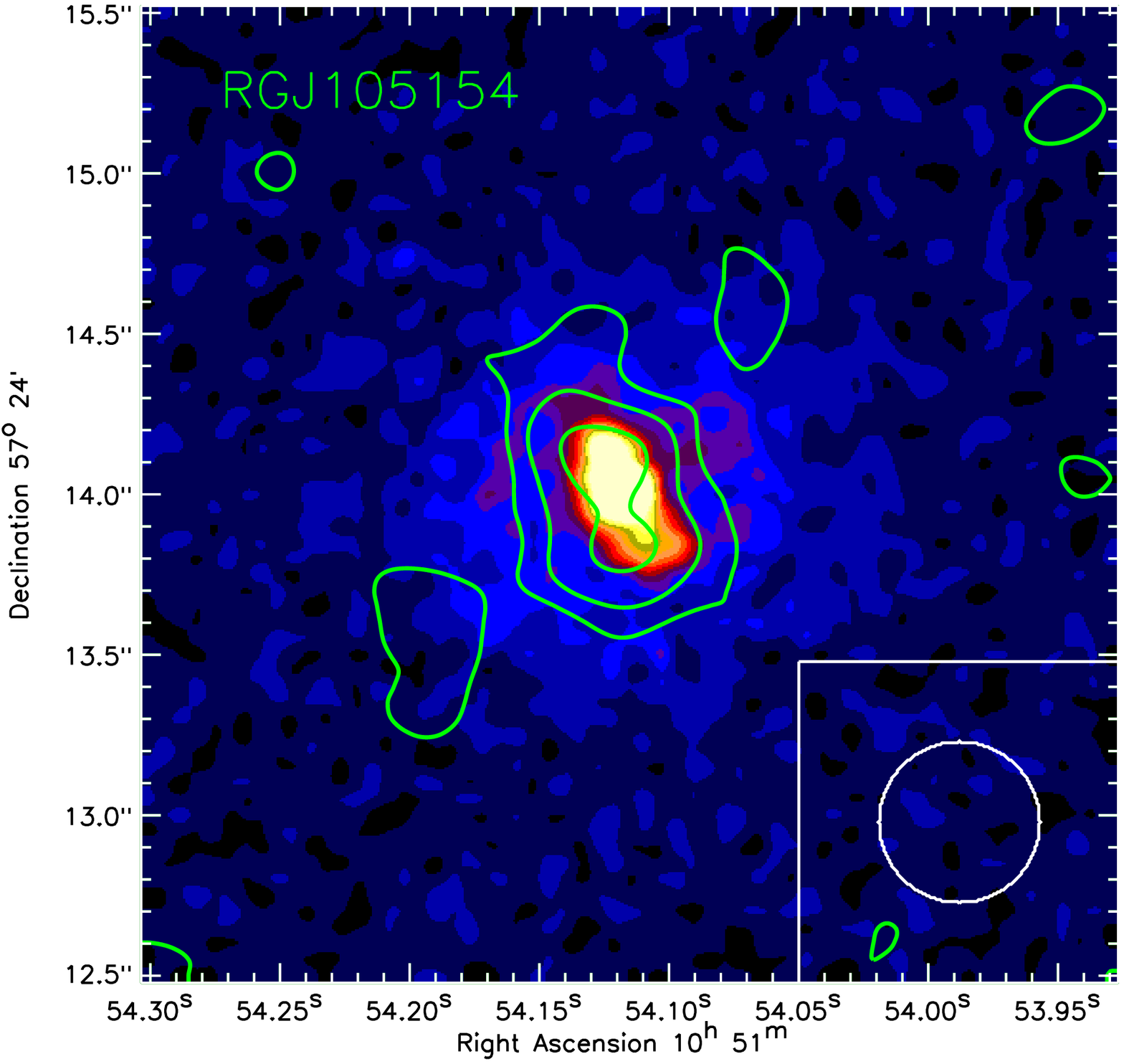}
\includegraphics[width=0.49\columnwidth]{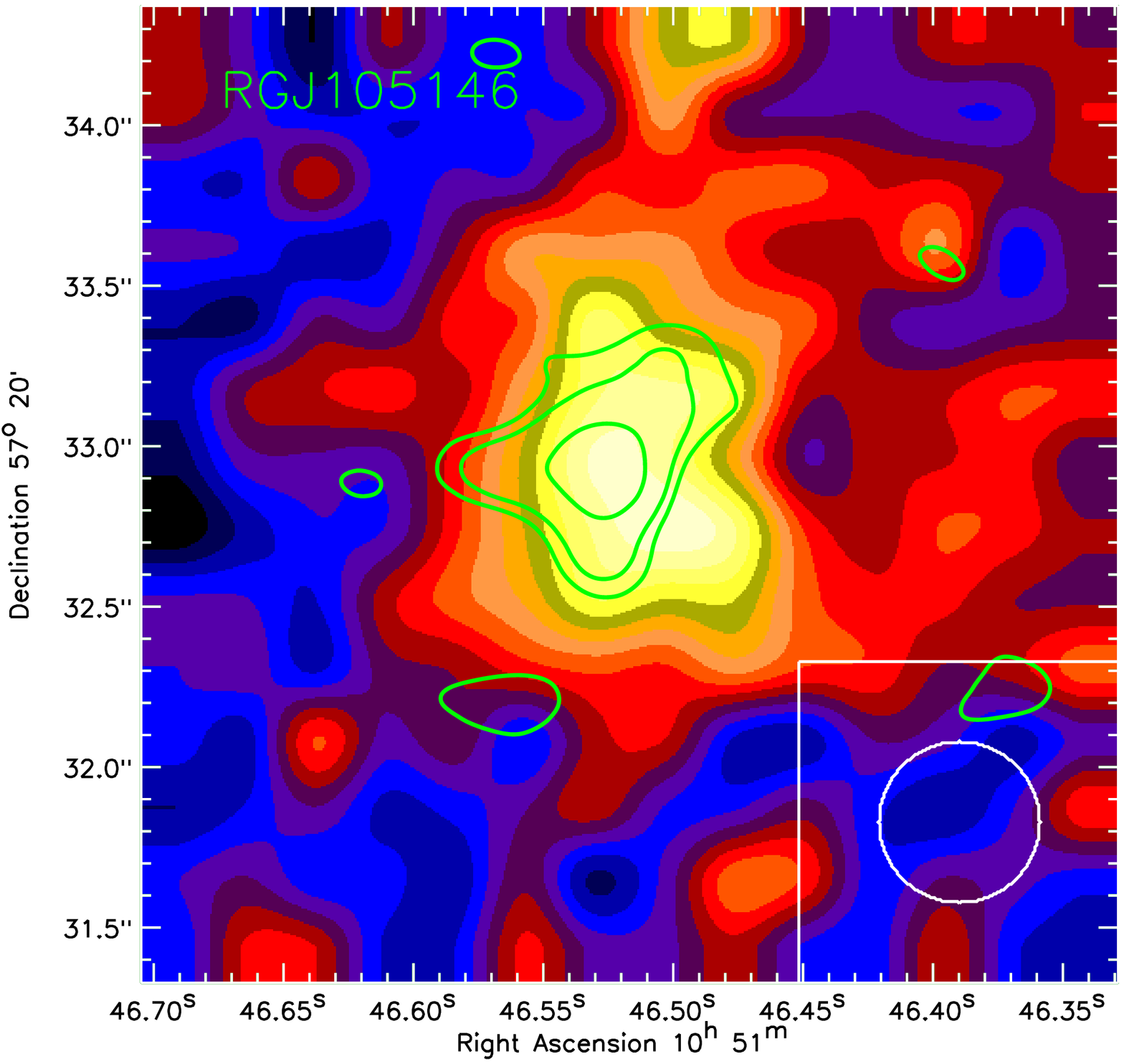}\\
\includegraphics[width=0.49\columnwidth]{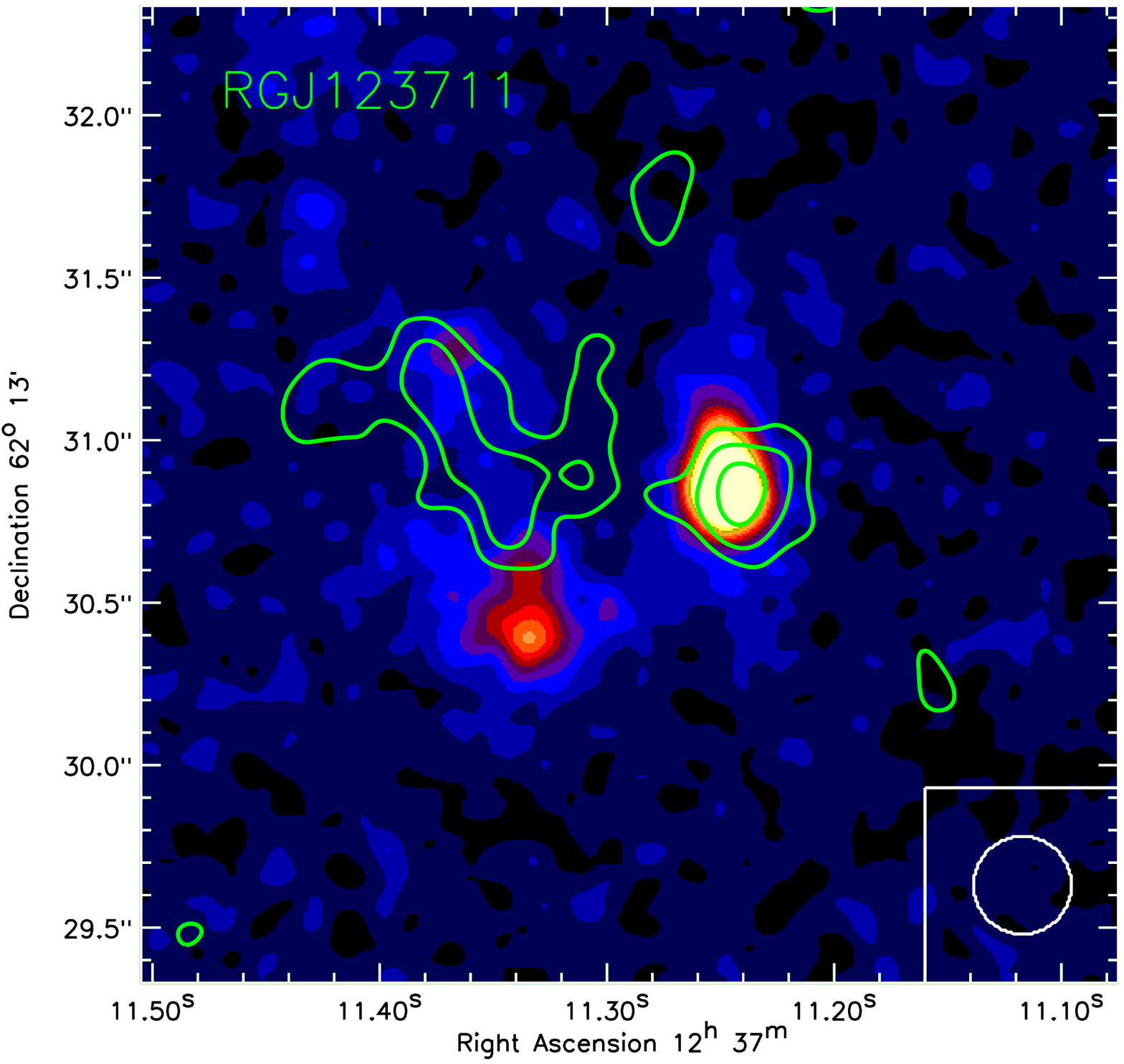}
\includegraphics[width=0.49\columnwidth]{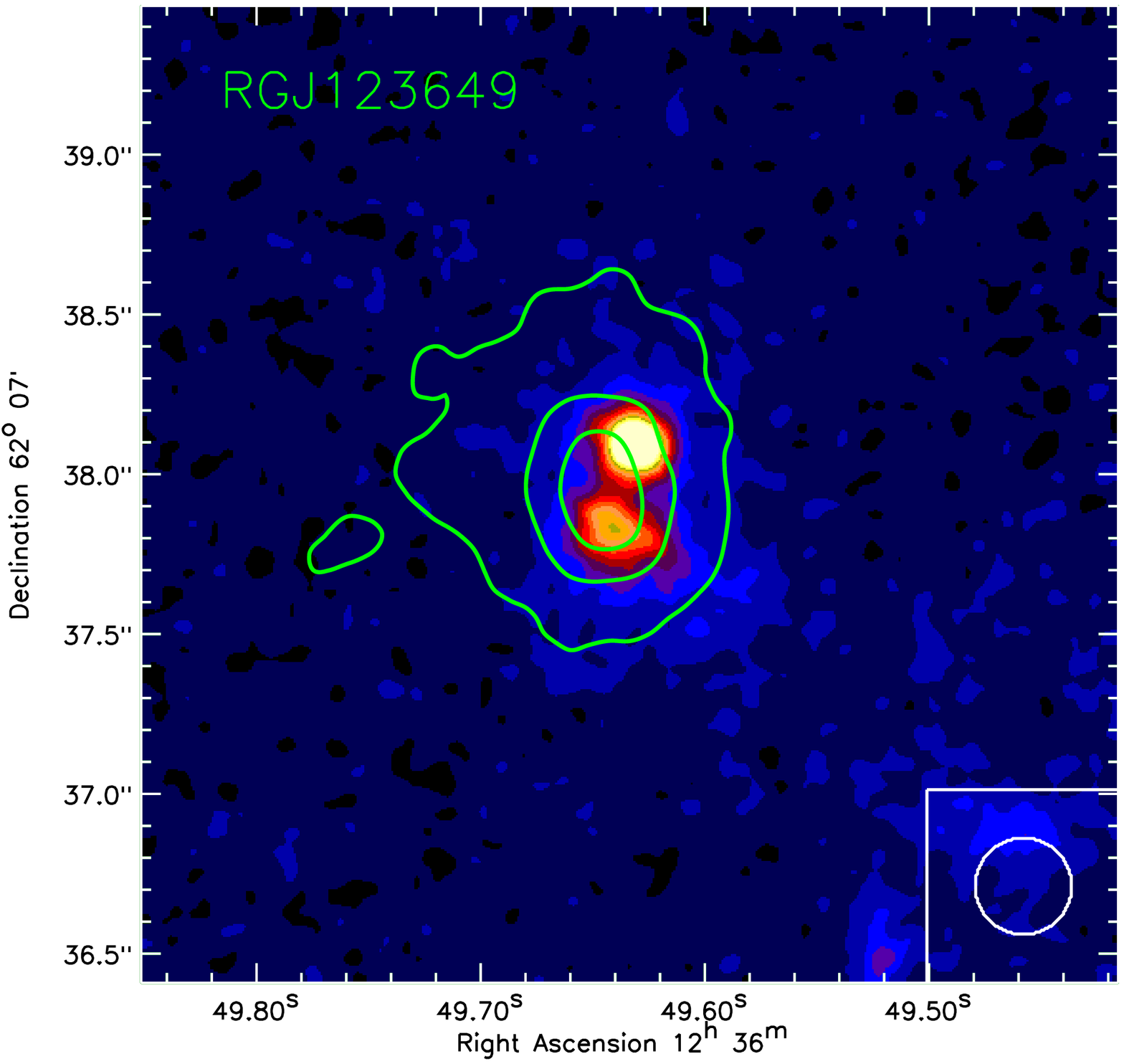}\\
\caption{
Contours of MERLIN high-resolution 0.3\arcsec\ radio emission overlayed on HST ACS $i$-band 
imaging for the eight hot-dust ULIRGs. ACS $i$-band imaging was unavailable for RG\,J105146 and
we used Subaru $i$-band in its place.  The unresolved MERLIN beam size FWHM is shown in the lower 
right of each panel in white.  The contour levels plotted are 40, 60, and
80\,percent of the peak flux density in each image.  Structure in radio emission is seen in each 
galaxy on 8\,kpc scales.}
\label{fig:MERLIN}
\end{figure}

We use the high-resolution MERLIN radio maps to assess the contribution of AGN to these galaxies 
by considering their radio morphology, which is shown as contour overlays on optical imaging in 
Fig.~\ref{fig:MERLIN}.  With resolutions of 0.3\arcsec\ and 0.5\arcsec\ per beam (for GOODS-N 
and the Lockman Hole respectively), the smallest resolvable structure at z$\sim$1.5 would be 
3-5\,kpc across; however, the typical size of an AGN emission region at radio wavelengths is much 
less than 1kpc.  This implies that an AGN dominated source would be completely unresolved in 
MERLIN radio maps \citep[e.g.][]{casey09a}.  Fig.~\ref{fig:MERLIN} shows that each of these galaxies has extended 
emission regions on 8\,kpc scales, suggestive of spatially distributed star formation, unlikely to 
be generated by AGN.  The effective radii corresponding to the surface areas of the radio emission 
regions ($>3\sigma$) are given in Table~1.  While the morphologies are irregular, we find that the 
effective radii average to 2.5$\pm$0.6\,kpc, which agrees with the size measurements of SMGs in 
\citep{biggs08a}.

\subsection{AGN Content}\label{sec:agn}

As their selection as SFRGs requires, these eight galaxies have rest-frame ultraviolet 
spectral features that are consistent with starbursts and do not exhibit features of 
active nuclear activity (e.g. no {\sc C\,IV} absorption).  Except for RG\,J123649, their 
X-ray luminosities (we assume $\Gamma=1.8$) are less than 3$\times$10$^{43}$erg\,s$^{-1}$.  
These low X-ray luminosities are consistent with star forming galaxies with little 
contribution from AGN \citep[e.g.][]{alexander05a}.  With the exception of RG\,J123649 
and RG\,J123711, the 6 remaining galaxies also have low 24\um\ flux densities ($S_{\rm 24}<250$\uJy). 
Low 24\um\ luminosity is more strongly associated with star formation than AGN emission.

The two hot-dust ULIRGs RG\,J123711 and RG\,J123649 which have the highest X-ray luminosities and mid-IR
luminosities of the sample might contain powerful AGN despite a lack of rest-UV AGN spectral features.  
The MERLIN radio emission in RG\,J123711 (bottom left panel in Fig.~\ref{fig:MERLIN}) is separated in two 
components: an extended region to the NE (63 percent of the integrated flux), and a bright compact region 
to the SW (37 percent).  The X-ray source coincides with this SW component.  In contrast, a large 
10$^{10}$\,M$_\odot$ molecular gas reservoir (a strong case that substantial star formation may be 
present) has been detected in CO and is morphologically separated evenly across the two NE and SW 
components \citep[][Casey et~al. 2009, in prep]{chapman08a}.  For these reasons, we believe the NE 
region is likely a star forming ULIRG merging with the SW region, an AGN-dominated galaxy.  

While the existence of an AGN in RG\,J123649 is supported by a bright X-ray luminosity and a compact core 
in optical imaging, there is no sign of AGN activity in the optical spectrum or in its $\sim$8\,kpc 
extended radio morphology.  We estimate that AGN emission is 40$\pm$30 percent and the star forming 
component is 60$\pm$30 percent of the radio flux by deconvolving an unresolved 0.3\arcsec\ PSF from 
its radio morphology.  We adjust the radio luminosity and inferred SFRs of both RG\,J123711 and 
RG\,J123649 to solely represent the estimated star forming components (see Table~1, ``RG\,J... SF 
component:'').  We have also labeled both galaxies as AGN in Fig.~1 and Fig.~2 to clarify that they 
differ from the other six.

It is possible that these two galaxies' 70\um\ flux densities are dominated by emission from 
torus dust surrounding the AGN (which can have very hot dust temperatures, $T_{D}\sim$1000\,K).  
However, disentangling the relative contributions of AGN and star formation to 70\um\ flux density 
is beyond the scope of this study since it requires detailed knowledge of the FIR SEDs.  For this 
reason, we exclude RG\,J123711 and RG\,J123649 from further analysis of the hot dust ULIRG aggregate 
properties, since we wish to characterize the properties of star formation dominated ULIRGs exclusively.

There is no evidence to suggest that the remaining six galaxies contain significantly luminous AGN.  With
a sample size of eight galaxies, two of which likely contain AGN, the total AGN fraction in the population
is 25 percent.  This is consistent with the estimated AGN fraction of Submillimetre Galaxies, between 20-40 percent,
from \citet{alexander05a}.

\subsection{Star Formation Rates}

We estimate SFRs from the VLA radio luminosities, using the radio/FIR correlation for 
star forming galaxies \citep{condon92a,helou85a,sanders96a}.  We abstain from calculating UV-inferred 
SFRs since they would be subject to significant (yet uncertain) extinction factors (as evidenced 
by the irregular MERLIN radio morphologies relative to the distribution of rest-UV flux in 
Fig.~\ref{fig:MERLIN}).  The FIR luminosities derived from the radio are consistent with the 
submm detection limits and the flux densities at 70\um\ as shown in Fig.~\ref{fig:all_seds}.  
The inner quartile (25-75 percent) of star formation rates in this sample is 200-1100\Mpy\ (both 
derived quantities, L$_{\rm FIR}$ and SFR are given in Table~1).

Star formation rate densities are found by dividing these SFRs by the surface areas of MERLIN 
radio emission regions.  Comparing these SFR densities to their theoretical maximum$-$the maximum
gas density divided by the local dynamical time \citep[see equation 5 of][ we use t$_{dyn}$=
4$\times$10$^{7}$yr]{elmegreen99a}$-$we can determine if the implied SFR density exceeds the theoretical
prediction.  While local ULIRGs with $\Sigma_{\rm SFR}\approx$200\,\Mpy\,kpc$^{-2}$ are forming stars at 
their theoretical maximum \citep[e.g.][]{tacconi06a}, none of the galaxies come within a factor of
four of exceeding their maximum SFR density limits.

We note that one of the eight galaxies of our sample, RG\,J123710, has been detected in CO gas by
\citet{daddi08a}.  The presence of a molecular gas reservoir $>$10$^{10}$\,M$_\odot$ indicates that 
high levels of star formation ($>$100\,\Mpy) can potentially occur.  In their interpretation of the 
detection, Daddi et~al. claims that RG\,J123710 is most likely a spiral galaxy undergoing modest-efficiency 
star formation lower than most SMG ULIRGs.  However, our detection of RG\,J123710 at 70\um\ directly confirms 
that it is a ULIRG.  More CO observations are needed to investigate if the rest of the hot-dust
ULIRG population exhibits modest star formation efficiencies or high, SMG-like efficiencies 
(Casey et~al., in preparation).

\subsection{70\um\ Undetected SFRGs}\label{sec:undetected}

Our eight galaxy sample likely represents the most luminous subsample of the hot-dust 
ULIRG population because of the current limitations in 70\um\ surveys.  To investigate
the possible lower-luminosity extension of the population, we analyze the 21 GOODS-N 
galaxies which were selected as SFRGs but not detected at 70\um\ (the Lockman sample
is excluded since its 70\um\ coverage has a much shallower depth).  Here, we stack 
70\um\ cutouts of these 21 SFRGs, centered on their VLA positions.  We use the stacking 
method described in \citet{huynh07a} and \citet{hainline09a} which was used to stack
$\sim$68 70\um\ undetected SMGs.  We have a detection of 0.45$\pm$0.15\,mJy in the SFRG
stacked image at 3.0$\sigma$, which is comparable to the 0.48$\pm$0.15\,mJy 3.2$\sigma$ 
detection for the SMGs of Hainline et~al.  This indicates that SMGs and 70\um-undetected 
SFRGs likely have similar 70\um\ luminosities, although larger samples of SFRGs are needed
to see if their relative detection rates at 70\um\ are similar.  Because of their detection 
at 70\um, the eight SFRGs of this paper are likely the highest luminosity (z$<$3) and
potentially the hottest dust temperature specimens of the SFRG population.

\section{Discussion}\label{sec:discussion}

\subsection{Volume Density}

The volume density is estimated by constraining the density of galaxies satisfying our selection 
criteria.  The density of these 70\um\ detected ULIRGs is 1.4$\times$10$^{-5}$\,Mpc$^{-3}$. 
However, the difference in volume density estimates between GOODS-N and Lockman Hole, which is
expected due to the 70\um\ sensitivity differences between the two fields, prevents us from
constraining the hot-dust ULIRG luminosity function well.

While this explicit population of hot-dust ULIRGs is $\sim5\times$ more rare than SMGs, there 
likely exists a larger population of $z>1$ ULIRGs with hot dust temperatures.  Firstly, 
spectroscopic incompleteness of $\sim$50 percent suggests the sample is likely twice as large
(although it is noted that SMGs are also affected by spectroscopic incompleteness).  
Secondly, there are likely additional ULIRGs at $z\sim2$ undetected in the radio (for instance 
from scatter in the FIR/radio relation).  Finally, the 70\um\ sensitivity limits of these 
maps do not exclude many of the other SFRGs in our parent sample from having warm enough dust 
to still represent hotter ULIRGs than SMGs.  As we have shown in our analysis of
the GOODS-N sample, proportionally more SFRGs are 70\um\ detected than SMGs; as also indicated
by the stacked result of 70\um\ undetected SFRGs, many more SFRGs might become 70\um\ detected if
the GOODS-N 70\um\ depth can be achieved in other fields.  The {\it Herschel Space Observatory} 
will dramatically improve the uniformity and depth of 70\um\ coverage (confusion limit of 0.04\,mJy 
at 2$\sigma$), potentially discovering many of the ULIRGs that lay just below the current MIPS 
detection limits.

\begin{figure}
\includegraphics[width=0.99\columnwidth]{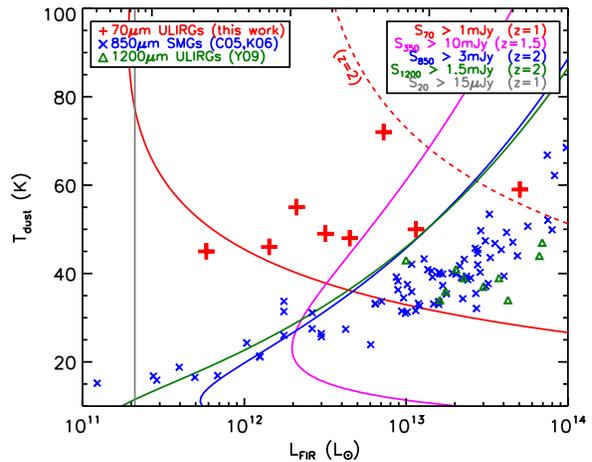}
\caption{
Far-IR luminosity against dust temperature illustrates the temperature/luminosity-biases of different
single-wavelength detection techniques, at 70\um\ (red), 350\um\ (magenta), 850\um\ (blue),
1200\um\ (green), and 20cm/1.4GHz radio (gray).  The boundaries on the plot are determined by flux density 
2$\sigma$ upper limits for these observations: 1\,mJy at 70\,\um, 10\,mJy at 350\,\um, 3\,mJy at 850\,\um, 
1.5\,mJy at 1200\,\um, and 10\,\uJy\ at 20\,cm (the limits for GOODS-N).  The boundaries also represent
a limit in redshift which is chosen based on the characteristic mean redshift of sources detected at that
wavelength.  Sources may be detected at the given wavelength on the high-luminosity side of each boundary.  
We overplot the SMG population which is selected at 850\um\ \citep{chapman05a,kovacs06a}, and z$\sim$2 
1200\um-luminous ULIRGs from the EGS field \citep{younger09a}, and the sample in this paper, detected 
at 70\um\ with a $z=1$ boundary (solid) and $z=2$ boundary (dashed).
}
\label{fig:lfir_td}
\end{figure}

\subsection{Selection Effect in $L_{\rm FIR}$-$T_d$ and Redshift}\label{sec:lfir_td}

\begin{figure}
\includegraphics[width=0.99\columnwidth]{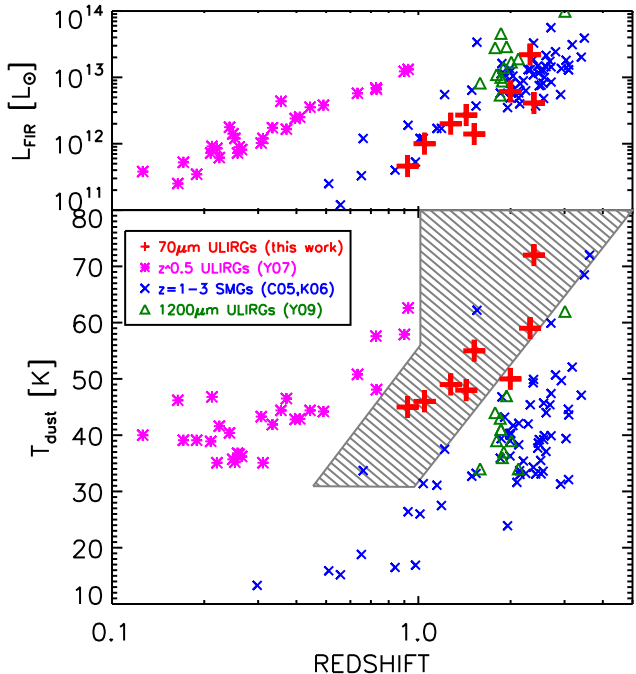}
\caption{
Far-IR luminosity and dust temperature as functions of redshift (log scale), with the same
symbols as in Fig.~\ref{fig:lfir_td}, with a low redshift 350\um\ galaxy population overplotted
for comparison \citep{yang07a}.  Above, the 70\um\ has similar luminosities to SMGs 
at the same redshifts.  Below we highlight the poorly explored region (shaded region) in 
dust temperature with redshift.  The 70\um-selected sample begins to sample this unexplored 
area.}
\label{fig:z_lfir_td}
\end{figure}

Fig.~\ref{fig:lfir_td} illustrates the selection effects of IR-wavelength observations as 
a function of FIR luminosity and characteristic dust temperature.  Assuming the radio/FIR 
correlation, radio detection is independent of dust temperature.  At $z=2$, a 10$^{13}$\,L$_\odot$ 
system is only detectable at 850\um\ or 1200\um\ if it has T$_d<40\,K$.  Since the mean redshift 
of radio-detected SMGs is $z=2.2$, the dust temperature selection-effect significantly impacts the proposed surface 
density of ULIRGs and thus their contribution to the global star formation rate density near $z\sim2$.  
If instead we select objects by the methods of this paper at 70\um, the dust temperature selection 
effect is inverted, and detection is possible for the hottest $z\sim$2 sources (T$_d>70\,K$).  
By lowering the redshift boundary from $z>2$ to $z>1$, detection is possible for 10$^{13}$\,L$_\odot$ 
galaxies that have T$_d>35\,K$.

While the FIR selection functions have hard boundaries in $L_{\rm FIR}$ and $T_{\rm d}$,
their dependence on redshift is important.  From Fig.~\ref{fig:lfir_td}, we see that
70\um\ selection is not ideal for $z>2$ samples where it is limited to temperatures 
T$_d$\simgt\,70\,K whereas at $z\sim1$ it can select a wider range of ULIRGs with 
T$_d$\simgt\,40\,K.  In Fig.~\ref{fig:z_lfir_td} we show the projections of both 
$L_{\rm FIR}$ and $T_{\rm d}$ with redshift.  This highlights that the luminosities 
of the 70\um-selected sample are comparable to SMGs at the same redshift.  However, 
as a function of dust temperature, the 70\um\ galaxies occupy an unexplored region 
in $z-T_{\rm d}$ which is heavily restricted by the selection biases illustrated in 
Fig.~\ref{fig:lfir_td}.

\subsection{Comparison of SED shape to SMGs}

\begin{figure}
\centering
\includegraphics[width=0.99\columnwidth]{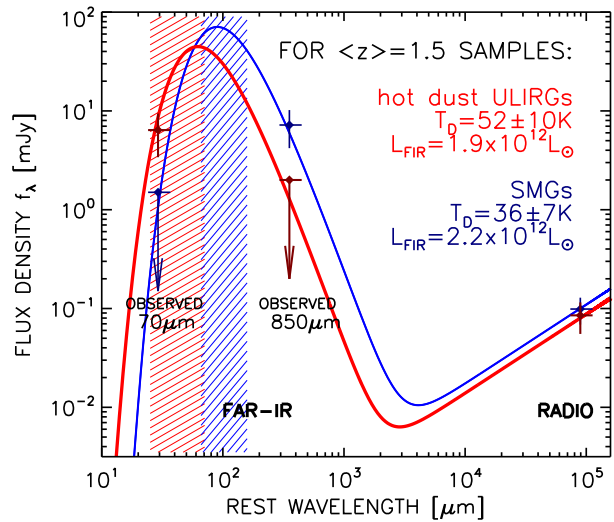}
\caption[SEDs of hot-dust ULIRGs relative to SMGs.]{
A composite spectrum of the six star forming, 70\um\ detected, hot dust ULIRGs relative to 
a composite SED for a sub-sample of SMGs with $<z>$=1.5.  We select several sub-samples of 
SMGs all with mean redshift 1.5 to mimic the redshift distribution of the hot-dust ULIRGs.  
The average flux densities for hot-dust ULIRGs are 6.4\,mJy at 70\um, $<$2\,mJy at 850\um\ and 50\uJy\ 
at 20\,cm.  The flux density points for SMGs are $<$1.5mJy at 70\um, 7\,mJy at 850\um\ and 
60\uJy\ at 20cm.  The red and blue vertical bands illustrate the range of peak fluxes 
corresponding to two temperature regimes; blue represents 36$\pm$7K and submm detected, 
while red represents 52$\pm$10K.  Note that at 350\um, both samples of 70\um\ and 850\um\ galaxies
would be equally easy to detect.
}
\label{fig:sed}
\end{figure}

A direct comparison of the observationally derived SED for these hot dust ULIRGs with SMGs is
shown in Fig.~\ref{fig:sed}.  We average the 70\um\ detections, 850\um\ upper limits, and 
radio detections of the six star formation dominant hot dust ULIRGs to form our composite 
SED; the composite SED adopts the mean redshift ($z=1.5$), FIR luminosity (1.9$\times$10$^{12}$\,L$_{\odot}$), 
and dust temperature (52\,K) of the six galaxy sample.  Since SMGs have a mean redshift of 2.2, 
we take several comparable sub-samples of GOODS-N and Lockman Hole SMGs to generate an average 
SMG SED with mean redshift 1.5.  The resulting mean luminosity of this SMG sub-sample is 
2.2$\times$10$^{12}$\,L$_{\odot}$ with dust temperature 36\,K.

Based on these SED fits, we do not expect SMGs to be detected at 70\um.  This agrees with 
observations; nearly all SMGs (68/73, $>$93 percent) are undetected $<$3$\sigma$ at 70\um\ 
\citep{hainline09a}. Those that are 70\um-detected are at low-$z$ and have higher luminosities 
than most SMGs at low redshifts.  To compare the SMG and SFRG populations, we compare their 70\um\ 
detection rates.  We limit this test to $z<2$ since the redshift 
selection functions for the populations differ \citep[submm+radio selection and radio selection 
have different redshift biases; see][]{chapman05a}.  This removes most effects of redshift-bias 
from radio detection in both samples.  SMGs and SFRGs have similar radio luminosities, so the 
same fraction of each should be detected at 70\um\ if they have the same distributions in dust 
temperature.  In GOODS-N, where the observations are deepest and the most complete, 4/8 (50$\pm$25 
percent) z$<$2 submm-faint radio galaxies are detected at 70\um\ but only 1/14 (7$\pm$7 percent) 
of submm-bright radio galaxies (SMGs) are detected \citep{chapman05a,pope06a,pope08a}.  The Lockman 
Hole observations do not have nearly the depth as those of GOODS-N and thus have fewer sources; 
however, the statistics are consistent with the finding in GOODS-N, with 3/7 (43$\pm$25 percent) 
submm-faint galaxies detected at 70\um\ and only 1/5 (20$\pm$20 percent) SMGs detected.  Therefore,
in contrast to SMGs, galaxies selected by our method (SFRG selection) are far more likely to be 
detected at 70\um\ (30-50 percent).

\subsection{Comparison to other IR Selection Techniques}\label{sec:comparison}
Since the advent of $Spitzer$ and other infrared observatories, the population of dusty 
galaxies studied at high redshift has grown substantially.  Many of these galaxies are 
more bolometrically luminous than typical UV-selected galaxies, but they may be poorly 
understood as a population due to faintness in the optical.  Furthermore, the optical 
counterparts for 850\um\ and 70\um\ sources are not easily identified due to the 
$\sim$12-20\arcsec\ beam size; for these reasons, substantial portions of SMGs and ULIRGs 
have no redshift identification.  Similar problems exist for other IR-selected galaxies.  
Here, we contrast the hot dust ULIRGs in this paper to other infrared-selected, dusty 
galaxy populations at $z\sim2$, noting potential overlap of selection techniques.

To circumvent the problems that arise from FIR selection (increased beam size and 
poor sensitivity limits), many dusty galaxies are selected in the mid-IR by their 
observed 24\um\ flux densities.  However, the selection of dusty star-forming LIRGs 
and ULIRGs through mid-IR 24\um\ continuum diagnostics is frequently contaminated 
by power law emission from AGN, and inferring star formation properties can require 
large correction factors to bolometric luminosity \citep[e.g.][]{dale05a,papovich05a}.  
All eight galaxies in our sample are classified as Dust Obscured Galaxies 
\citep[DOGs;][]{dey08a,pope08a}; however, the selection criteria of DOGs is so broad 
that without detailed additional information (the radio maps, FIR detections and 
non-detections, and near-IR photometry) it is difficult to understand the importance 
or evolutionary significance of this classification.  We further emphasize that 3/8 
specimens in our 70\um\ sample have S$_{24}<$300\uJy\ (and are only selected as DOGs 
under the selection described by Pope et~al.), revealing that ultraluminous activity 
at $z=$1-3 can be missed by 24\um\ selection criteria.

A recent study by \citet{younger09a} presents 12 $z\sim2$ ULIRGs in the EGS 
field, selected in the $Spitzer$ IRAC and MIPS bands and followed up with MAMBO 
at 1200\um\ (where 9/12 are detected).  They have a characteristic dust temperature 
$41\pm5\,K$ (using $\beta=1.5$ modified blackbody fits) and are shown in 
Fig.~\ref{fig:lfir_td} and Fig.~\ref{fig:z_lfir_td}. The 12 Younger~et~al. 
galaxies bridge the gap between SMGs and hot 70\um-detected sources in temperature but
are much more luminous than the 70\um\ sample. They have a mean 1200\um\ flux density
of $\sim1.6$\,mJy where our 70\um\ sample is uniformly undetected\footnote{The depth of 
MAMBO coverage in GOODS-N is $\sim$0.7-0.8\,mJy RMS compared to the $\sim$0.4\,mJy
photometric RMS of the EGS pointings, while Lockman Hole has 0.8-1.0\,mJy RMS.}.  This 
would imply that the 12 EGS galaxies have 850\um\ flux densities $\sim$3-6\,mJy, 
FIR luminosities (8-1000\um) $>$10$^{13}$\,L$_\odot$ and thus could be classified 
as SMGs.  The selection of the Younger et~al. sample in the mid-IR, by the presence
of a prominent stellar bump in $IRAC$ photometry and dust at 24\um, is similar to
the DOG selection, although with a few additional constraints reveals a potentially 
strong probe of ultraluminous activity at high redshift.  However, more far-IR 
observations of $z\sim2$ mid-IR selected samples are needed to assess the selection's
completeness in choosing both star forming ULIRGs and AGN dominated ULIRGs at $z>1$.

Selection at 350\um\ at z$\sim$2 has similar temperature biases as 850\um\ and 1200\um\ 
due to its high flux limit (S$_{350}>$10\,mJy at 2$\sigma$).  As seen in 
Fig.~\ref{fig:sed}, 350\um\ selection at $z\sim1.5$ has the potential to select SMGs and 
hot-dust ULIRGs equally well since it samples the dust SED near its peak for a wide range 
of temperatures.  However, until the depth of 350\um\ imaging can be improved, current 
observing facilities allow only 350\um\ detections of the most luminous or low redshift 
sources \citep[e.g.][]{yang07a}.  The Balloon-borne Large Aperture Submillimeter
Telescope (BLAST) has recently mapped the Chandra Deep Field South at 250\um, 350\um, 
and 500\um\ at greater depth than has been done before in any field at those wavelengths 
\citep{patanchon09a}.  While useful for measuring the flux densities of already known 
ULIRGs and high-redshift galaxies, the statistically significant source counts in the 
BLAST map are limited \citep[with only $\sim$12 galaxies detected at z$>$1;][]{viero09a}.  
When it begins full operations, the {\it Herschel Space Observatory} will be another 
observing tool at 350\um\ and will dramatically improve the depth of 70\um\ observations 
in many heavily observed fields, thus providing better statistics at shorter FIR wavelengths.  
In addition, future work from the {\sc SCUBA2} instrument at 450\um\ have the potential to 
expand the sample of $z\sim1-3$ ULIRGs through deeper observations near the peak of the 
dusts' SED.

\section{Conclusions}

This paper has observationally demonstrated that submillimetre wavelength observations of galaxies 
at $z\sim1-3$ have a strong temperature bias which preferentially detects galaxies with 
cooler dust temperatures.  By selecting ULIRGs that are detected at 70\um\ (and verifying that they are 
starburst dominated) we have found a set of galaxies which are undetected in the submillimetre
yet still represent some of the most luminous systems at $z>1$.  Their volume density is 
1.4$\times$10$^{-5}$\,Mpc$^{-3}$, approximately 5$\times$ more rare than SMGs, although
this has likely been underestimated due to spectroscopic incompleteness and large variations in
the sensitivity limits of 70\um\ surveys.

Like most IR-selected dusty galaxies, the SFRs in these hot dust ULIRGs are on average much higher 
than the majority of UV-selected star-forming galaxies at the same epoch.  While some UV-selected 
sources have high stellar masses \citep[][ Hainline~et~al., in preparation]{shapley05a} they show 
consistently lower SFRs, reiterating that the UV misses the most dramatic star-forming galaxies at 
high redshift.  Due to their similarity with SMGs (in terms of stellar mass, UV-spectra, FIR and 
radio luminosities), the hot-dust ULIRGs represent an extension of SMGs towards a wider range of 
dust temperatures at z$\sim1-3$, similar to the already observed wide range in dust temperatures 
seen in local ULIRGs. 

We have shown that AGN contribute little to the FIR luminosities in six of the eight hot dust ULIRGs. 
The detection of extended radio emission (with typical radii 2-3\,kpc) suggests spatially 
distributed star formation rather than compact AGN. In addition, the UV-spectral properties and low 
X-ray luminosities are consistent with star formation.  The two ULIRGs that do show signs of AGN activity 
also have evidence for substantial star formation and are therefore comparable to SMGs which contain bright 
AGN.

Combining our hot-dust ULIRG population with the SMGs results in a sample of $z\sim2$ ULIRGs with less 
temperature-dependent bias than the SMG population alone. However, we have shown that searches for high 
redshift FIR luminous galaxies are likely still somewhat incomplete.  Even at $z\sim2$, there are likely 
other FIR luminous galaxies which are not well characterized by current submm or FIR observations. 
$Herschel$ and {\sc SCUBA2} will help push detection of more $z>1$ ULIRGs in the 50-500\um\ range
while more work with deeper radio observations (from {\it e}-MERLIN and EVLA) will be needed to 
find potential ULIRG activity beyond $z>3$.  By working towards completeness in the $z>1$ ULIRG
population, we will learn about the role of heavy, short-lived star formation in the formation
and evolution of galaxies in the early Universe.

\section*{Acknowledgments}
We thank the anonymous referee for helpful comments which improved the paper.
This work is based, in part, on observations made with MERLIN, a National Facility operated by 
the University of Manchester at Jodrell Bank Observatory on behalf of STFC, and the VLA of the 
National Radio Astronomy Observatory, a facility of the National Science Foundation operated 
under cooperative agreement by Associated Universities, Inc. CMC thanks the Gates-Cambridge 
Trust, and IRS thanks STFC for support.

\label{lastpage}

\end{document}